\documentclass[12pt,preprint]{aastex}

\slugcomment{A Search for SS Cyg Outburst Predictors}

\shorttitle{Photometry and Trend Analysis of SS Cyg Outbursts}
\shortauthors{Price, Henden, Foster, Petriew, et al.}

\begin{document}


\title{SS Cyg Outburst Predictors and Long Term Quasi-Periodic Behavior}


\author{A. Price\altaffilmark{1,2}, A. A. Henden \altaffilmark{1}, G. Foster \altaffilmark{1}, V. Petriew \altaffilmark{1}, R. Huziak 
\altaffilmark{1}, R. 
James \altaffilmark{1}, M. D. Koppelman \altaffilmark{3}, J. Blackwell \altaffilmark{1}, D. Boyd \altaffilmark{4}, S. Brady \altaffilmark{1}, Lewis M. Cook \altaffilmark{1}, T. Crawford 
\altaffilmark{1}, B. Dillon \altaffilmark{1}, B. L. Gary \altaffilmark{5}, B. Goff \altaffilmark{1}, K. Graham \altaffilmark{1}, K. Holland \altaffilmark{1}, J. Jones \altaffilmark{1}, R. Miles \altaffilmark{1}, 
D. Starkey \altaffilmark{1}, S. Robinson \altaffilmark{1}, T. Vanmunster \altaffilmark{1}, G. Walker \altaffilmark{1} }

\affil{\altaffilmark{1}American Association of Variable Star Observers
    Clinton B. Ford Astronomical Data and Research Center, 
    49 Bay State Road,
    Cambridge, MA 02138 USA
}
\affil{\altaffilmark{2}Wright Center for Science Education,
    Tufts University,
    4 Colby St.,
    Medford, MA 02155 USA}

\affil{\altaffilmark{3}University of Minnesota, 
       Astronomy Department,
       Minneapolis, MN, 55455  USA}

\affil{\altaffilmark{4}British Astronomical Association Variable Star Section,
West Challow, OX12 9TX, UK}

\affil{\altaffilmark{5}G95 Hereford Arizona Observatory,
       Hereford, AZ USA}

%%\affil{National Optical Astronomy Observatories, Tucson, AZ 85719}


%%\altaffiltext{1}{Patron, Alonso's Bar and Grill}


\begin{abstract}
We report null results on a two year photometric search for outburst predictors in SS Cyg.
Observations in Johnson V and Cousins I were obtained almost daily for multiple hours per night 
for two observing seasons. The accumulated data are put through 
various statistical and visual analysis techniques but fails to detect any outburst predictors.
However, analysis of 102 years of AAVSO archival visual data led to the detection of a correlation between a long term 
quasi-periodic feature at around 1,000-2,000 days in length and an increase in outburst 
rate.

\end{abstract}



%%\keywords{variable stars: cataclysmic --- variable stars: individual{SS CYG}}



\section{Introduction \& Prior Work}

     SS Cyg is one of the most popular and studied variable stars. It is a cataclysmic variable (CV-a.k.a 
dwarf novae) of the U Gem (UG) class, and prototype of the SS Cyg (UGSS) subclass \citep{kho85}.  UG class
CVs are a binary system consisting of a late main sequence secondary star orbiting a white dwarf primary. The
secondary star continuously fills its Roche lobe and transfers mass to the primary, which forms an accretion
disc around the primary as angular momentum is conserved and spread within the disc. Occasionally the    
accretion disc flares up into a bright state referred to as an "outburst". The UGSS subclass have outbursts
that are usually separated on the order of months, but are largely unpredictable. Outbursts typically have an
amplitude of 2-6 magnitudes and last a few days to a few weeks.

    The most popular fundamental model for the outburst mechanism in UG stars is the disk instability model  
first proposed by Osaki (1976). In it, the region where the in-falling matter hits the accretion disc, known
as the hot spot, reaches a critical mass. This is a result of a higher rate of mass transfer from the
secondary to the disc than from the disc to the white dwarf (due to viscosity).  When the 
leftover material reaches a critical mass, the accretion disc becomes unstable. Viscosity increases, along with the mass transport rate,    
through the disc. Both increases are driven by angular momentum. Eventually the system blazes in additional
light as the mass built up emits energy due to enhanced ionization of Hydrogen. Only about 3-10\% of the 
material from the disc actually falls onto the white dwarf (see panel 3 of figure 13 in Cannizzo [1993]).

\subsection{Outburst Types}

      SS Cyg's quiescence magnitude is around v=12. Outbursts are usually expected every 4-10 weeks,         
with a duration of 7-18 days each \citep{jev03}. During outburst, the star reaches 
maximum light near magnitude v=8. Cannizzo and Mattei (1992) describe the outbursts as bimodal with long
outbursts typically lasting $>$12 days and short outbursts lasting $<$12 days.  They assigned the letter "L" to the
long outbursts and "S" to the short and anomalous outbursts. The anomalous outbursts tend to have a
linear, lower rate of change and smaller amplitude than long and short outbursts (Figure 1). They found the most common sequence as LS (with
134 occurrences), LLS (69), LSSS (14), and LLSS (8). About 89\% of all outbursts can fall into one of these
four sequences. 

\begin{figure}
\epsscale{1}
\plotone{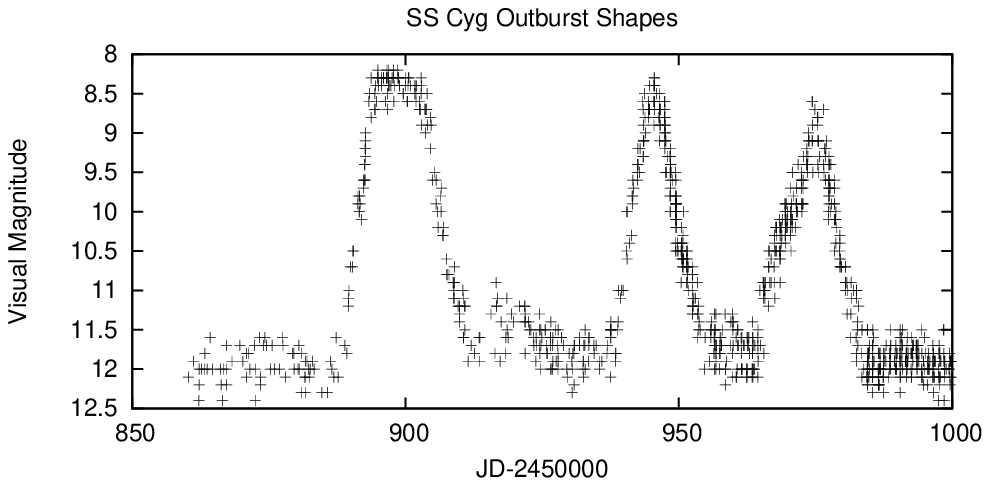}
\caption{A sample of the SS Cyg visual light curve showing the three category of outbursts: long, short and anomalous (left to right).}
\end{figure}

\subsection{Outburst Cycles \& Periodicity}

    Honey et al. (1989) photometrically observed SS Cyg over a single long and a single short outburst cycle.
They reported substantial flickering on the order of minutes but were unable to find any modulation (including near the orbital 
period) or periodicity to an amplitude limit of 0.05 magnitude. Giovannelli, Martinez-Pais and Graziati (1992) review the 
flickering behavior of SS Cyg in both outburst and quiescence, but find no correlation with outburst type, although 
they do find an inverse correlation between flickering amplitude and system brightness.

     Cannizzo and Mattei (1992) analyzed the historical light curve of the American Association of Variable Star
Observer's (AAVSO) International Database (hereafter: AID), which at the time included 29,387 individual    
daily means from 1896 September 27 to 1992 April 7.  They found no correlation between outburst duration time 
and cycle time but did confirm a correlation between quiescent magnitude and cycle time, caused by a variation 
in the mass transfer rate of up to a factor of 2 on yearly time scales and a variation of 20\%-30\% on decadal 
time scales. They did not find any periodicity in cycle time.

      In a later study \citep{can98}, using a smaller subset of the data (1963 - 1997), they discovered a variable decay rate in 
some outburst declines. In a minority of the outbursts, the decay rate slows down about two-thirds of the way into the decline for about a day and 
a half, then the decay returns to its previous rate.  The significance of the break is proportional to the length of the total 
decline. The break manifests itself as a 20\%-300\% decrease in the decay rate for around one day.  Cannizzo and Mattei (1998) refers to this 
phenomenon as a "glitch", we refer to it hereafter as the "Cannizzo Glitch".

     Ak et al. (2001) analyzed the quiescence magnitude and outburst cycles of 23 dwarf novae mostly using
Fourier analysis. They found no correlation between outburst cycle period and masses of component stars, mean
outburst interval, mean outburst duration, mean decline and rise rates of outbursts, absolute quiescent   
magnitudes and outburst states.

    In a poster, Hill and Waagen (2005) suggested the existence of a pre-outburst brightening of about 2\% over the course of 
around 12 hours followed by a plateau which lasts another 12 hours leading up to the onset of the outburst. The analyzed data 
set included about half of the AID visual data and excluded CCD observations. This was the only search for predictors found in 
published literature.

\subsection{Long Term Variation}

    Kiplinger, et al. (1988) report a 0.15 magnitude amplitude, 7.2 year period in the quiescent behavior of SS Cyg, determined 
through Fourier analysis of one day means in the AID data from 1896 - 1984. Also, Bianchini (1988) discovered a 
similar 6.9 year period through Fourier analysis of the intervals between outbursts and attributes it to solar type 
variation in the secondary \citep{Bia92}.

    However, Richman, Applegate and Patterson's (1994) own analysis of the individual data points plus various long term moving averages find that the power of the 
reported $\sim$7 year period is not significantly greater than that of other low frequency signals in the power spectrum and 
caution that the reported significance is due to the eye being, "...very prone to spot one to three cycles of periodic behavior 
in any randomly varying time series." However, their analysis excluded all quiescent data points within 12 days on either side 
of the outburst, thus if the source for the reported $\sim$7 year period was found in activity near the outburst then it would 
be hidden from their analysis. 

    Hemplemann and Kurths (1990) O-C analysis found secular variation within 100 years as well as deviations from the sample mean over intervals 
of a few tens of cycles. Jevtic, 
Mattei and Schweitzer (2003) performed a nonlinear analysis using Poincar\'{e} section and found a period of around 52 years. It 
is unclear whether the two periods are related as a fundamental and a harmonic pair.

\section{Observations}

     SS Cyg was observed in the {\it V} and Cousins {\it I} (hereafter {\it I}$_{c}$) \citep{Cou76} bands over 483 nights 
during a period from May 1, 2005 to February 5, 2007 by AAVSO observers participating in a coordinated campaign. The average length of coverage 
per observing day was 446 minutes 
in {\it V} and 348 minutes in {\it I$_{c}$}.  In total, 108,612 individual observations in {\it V} and 14,292 observations in 
{\it I$_{c}$} were obtained from 24 observing locations during the course of the campaign.

   Each observer reduced his/her own differential photometry using comparison stars published on an AAVSO chart. The values from 
those stars were determined via precision photometry obtained over multiple nights using the U.S. Naval Observatory - Flagstaff 
Station (NOFS) 1.0m telescope along with a large set of Landolt standards \citep{lan92} having a range of color and airmass.  
The raw data and reported uncertainties are available for download from the AAVSO web site.

   Uncertainty in the individual observations varied widely. When simultaneous data was available, the data with the 
higher reported uncertainty was excluded. 
A collective uncertainty for the entire dataset was determined by averaging 
the observations into one day bins and then adding the standard deviation in quadrature. The result gives an zeropoint 
uncertainty of 0.1 magnitudes in {\it V} and 0.16 magnitude in {\it I$_{c}$}. This includes both stochastic uncertainty and 
uncertainty caused by the flickering.  When considered separately, most individual datasets had a stochastic 
uncertainty (precision) between 0.01-0.05 magnitudes.

   In addition to our photometric data, we also analysed 366,614 visual observations available from the AAVSO International 
Database \citep{Tur06}. The observations date from May 17, 1904 to July 31, 2006. Previous studies have determined an 
uncertainty of 0.2-0.3 magnitudes for visual observations while exceptional visual observers can reach a precision of 0.02 
magnitudes \citep{pri06}. The visual bandpass varies by observer, but in general is slightly bluer than {\it V} \citep{Col99}.

\section{Analysis}
\subsection{Photometric Data}

   The resulting light curve covers most of the 2005 and 2006 observing seasons (Figure 2). It includes four long outbursts, six short outbursts, and no anomalous 
outbursts. Inspection of the AAVSO visual light curve confirms no outbursts were missed.

    There was considerable flickering in both bands during quiescence (Figure 3). A straight line was removed from a dataset consisting of ({\it 
V}-{\it I$_{c}$}) measurements converted to flux space \citep{bes79}. No trends were found in either the standard deviation of the coefficients or the residuals of the straight line.
Note that due to the flickering in this star, any color not formed from simultaneous observations has built-in error greater than the Poisson
or measurement error.

    To look for predictors, we folded the {\it V}, {\it I$_{c}$}, ({\it V}-{\it I$_{c}$}) and visual datasets onto a single light curve correlated 
with the onset of outburst, defined as the first observation greater than 0.1 Jy in {\it V}, 0.14 Jy in {\it I$_{c}$} and 0.25 Jy for visual 
observations (Figure 4) and which is greater than 20 days since the start of the previous outburst (to filter out observational scatter around the 
outburst thresholds during the decay). We did not find any consistent predictor in the light curves. We also folded the residuals of the detrended 
({\it V}-{\it I$_{c}$}) dataset to look for any change in the flickering rate prior to outburst, but none were found.

    A Fourier analysis \citep{fer81} of the quiescent observations is dominated by red noise caused by the flickering (Figure 5). The orbital period was detected 
at 0.275300 +/- 0.000006 days along with a strong harmonic at 0.13757 +/- 0.000001 days.  No other periodic signals were 
detected. Significance in all Fourier analysis was determined at the 3 sigma level by tripling the standard deviation from the average power of each spectrum computed with the minimum frequency resolution supported by the data. 

    The Cannizzo Glitch was originally detected in around 1/6 of the outbursts of SS Cyg. \citep{can98} We did not detect the glitch among our ten analyzed outbursts. 

\subsection{Visual Data}

    The AID collection of visual observations was analysed to look for periodicity in the outbursts and quiescence activity which could be a predictor of future outburst 
behavior. One day means for the complete 102 year visual database were computed in magnitude space so errors will remain unbiased in the magnitude domain rather than  
becoming biased after transformation into flux space. To look for long term periodicity, such as Kiplinger's 7.2 year and Bianchini's 6.9 year periods (hereafter 
{\it KB's}) periods, a Fourier analysis for signals in the 1,000-10,000 day range was computed (Figure 6). It reveals multiple signals, including KB's periods, but 
nothing that stands out on its own above background activity levels. This confirms the results of Richman, Applegate and Patterson's earlier analysis using a smaller 
sample of the visual database.

However, periods consistent with the KB reports are present from the beginning of the light curve until 
approximately 2426500. Fourier analysis with the CLEANEST algorithm \citep{fos96b} of data until 2426500 reveals 
a strong signal at 3076 days (Figure 8). Another strong signal appears around 2436500, this one around 1,530 
+/- 94 days. It only lasts roughly 13,600 days, but it also appears roughly 10,000 days earlier and again 10,000 days 
later.

    A weighted wavelet Z-transform \citep{fos96a} (Figure 7) of the same frequency range with a window size 
of two full cycles between the inflection points of the Gaussian envelope (c=0.0125) shows intermittent presence of 
various signals.  The WWZ is designed to be approximately an F-statistic with non-integer numbers of degrees of freedom.  But with a generous quantity of data, it 
can be treated roughly as chi-square with three degrees of freedom.  Hence the values of its peaks are statistically significant, although we express two caveats.  First, a significant 
value does {\it not} indicate that the data matches a periodic or pseudo-periodic fluctuation, only that it contradicts the null hypothesis (that there is no signal at all, just noise).  
Second, the stated statistical behavior is for a white-noise process; for red noise the "critical values" will be higher.  Nonetheless, the peaks in the time-frequency plot are at high enough 
values that we consider them strong evidence of transient pseudoperiodic fluctuation.

  We also calculated the time between onset of outbursts in the visual dataset. Due to inconsistent coverage, we 
ignored the interval 2417586 - 2419339 because we could not be sure an outburst was not missed by the 
observers. Simply ignoring the gaps during this interval, as opposed to the entire interval, would have created a 
selection effect that favored shorter times between outbursts. We found 703 outbursts with an average outburst onset 
interval of 49.2 +/- 15.8 days.

  To look for changes in this interval, we averaged the outburst intervals in 500d bins. The result (Figure 9) shows a 
possible relationship between outburst interval and the 1,000 - 2,000d periods detected in the wavelet analysis. We 
computed the peak periods via Fourier analysis of 10,000d windows centered on the midpoints of each of the 500d mean bins. 
We dropped the first and last 10 bins since they had $<$10,000d of data in them. Finally, we computed a bivariate 
Pearson Correlation coefficient \citep{tur08} between the peak period and the average outburst intervals. The result 
was a coefficient of -.356, significant to the 0.01 level, which is moderately strong according to Cohen's guidelines. 
Thus we conclude there is a moderate negative relationship between the strength of 1,000-2,000d periods in the 
quiescent visual database and the intervals between outbursts.

\section{Conclusion}

     We have analysed intensive {\it V} and {\it I$_{c}$} band observations of SS Cyg over two observing seasons. We were unable to detect any 
predictors or other activity which could be used to predict an oncoming outburst. However, the combined uncertainty of our photometric data was 
high enough to warrant further investigation with more precise observations. In particular, some of the data suggest a rise in the ({\it 
V-I$_{c}$}) color beginning five days prior to outburst. The rise level was barely within our uncertainty so we cannot report it. A similar 
dataset in ({\it B-$I_{c}$}) may increase the photometric sensitivity to such a feature, if it exists.

     We also analysed 102 years of AAVSO visual observations. No periodicity was detected which had not already been reported. However, long term 
quasiperiodic features were detected on the order of 1,000-2,000 days. Their existence is moderately correlated with shortened intervals between 
outbursts. If the quasiperiodic features are due to solar type variation, as previously reported, then it is possible that the 
enhanced activity drives additional mass transfer thus decreasing the mean time between outbursts.

\section{Acknowledgments}

  We would like to thank Eric Chaisson and Matthew Templeton for advice in preparation of this manuscript and Dave Tandy for support during 
the observing campaign.

%%\begin{deluxetable}{ccccccccccc}
%%\tabletypesize{\footnotesize}
%%\rotate
%%\tablecolumns{2}
%%\tablewidth{0pc}
%%\tablecaption{2000-3000d quiescence quasiperiodicity in visual observations from 1902-2006}
%%\tablehead{             
%%        \colhead{{\it Time span (JD)}} & \colhead{{\it Quasiperiod(s) }}  }
%%\startdata
%%2416617 - 2426617 &  2971.997836$\pm$71.794965  \\ 
%%2426617 - 2436617 &  2074.000332$\pm$76.708974  \\
%%2436617 - 2446617 &  2651.999342$\pm$58.711582  \\
%%2446617 - 2453947 &  2708.999296$\pm$N/A  \\
%%\enddata
%%\end{deluxetable}

\begin{figure}
\epsscale{1}
\plotone{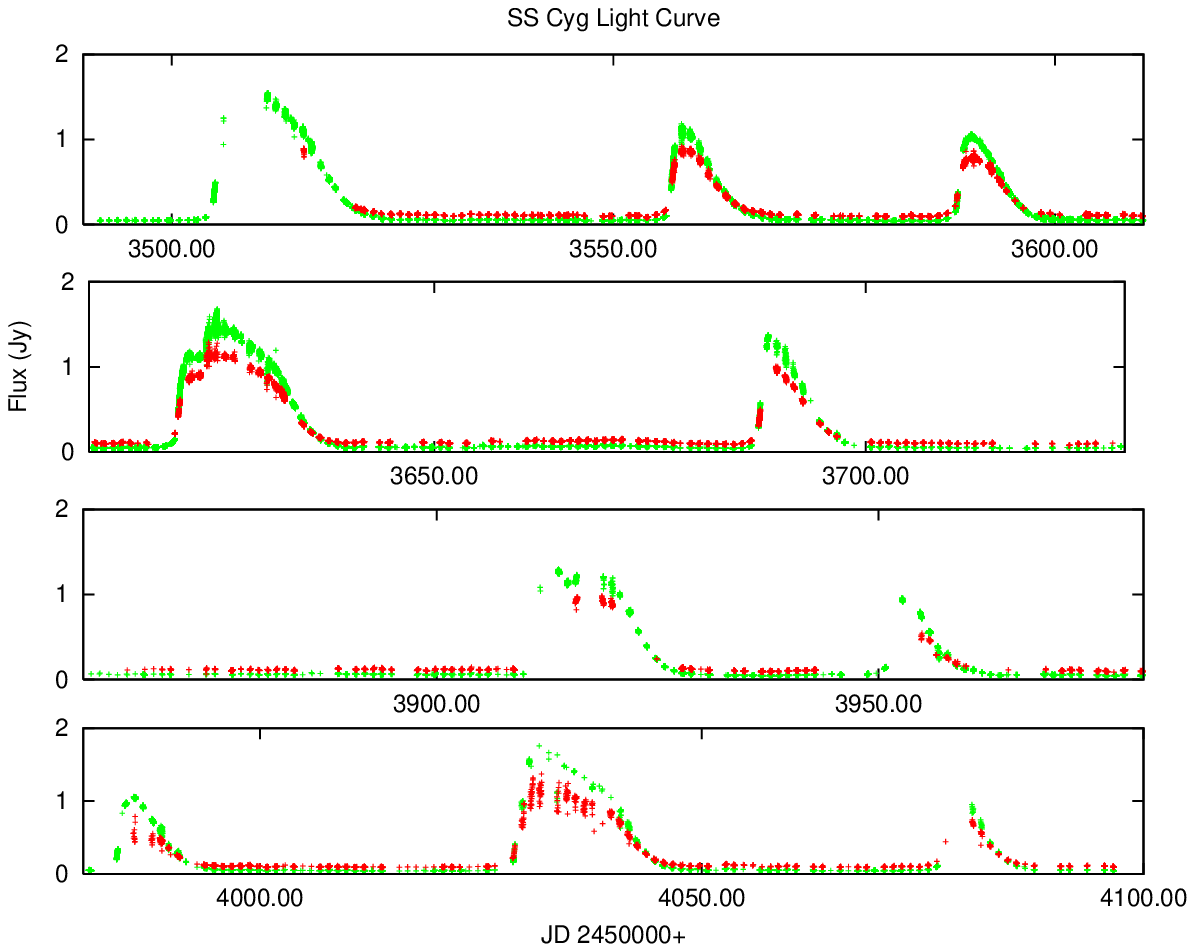}
\caption{{\it V} (green) and {\it I$_{c}$} (red) band photometric light curve for the 2005 and 2006 observing seasons.}
\end{figure}

\begin{figure}
\epsscale{1}
\plotone{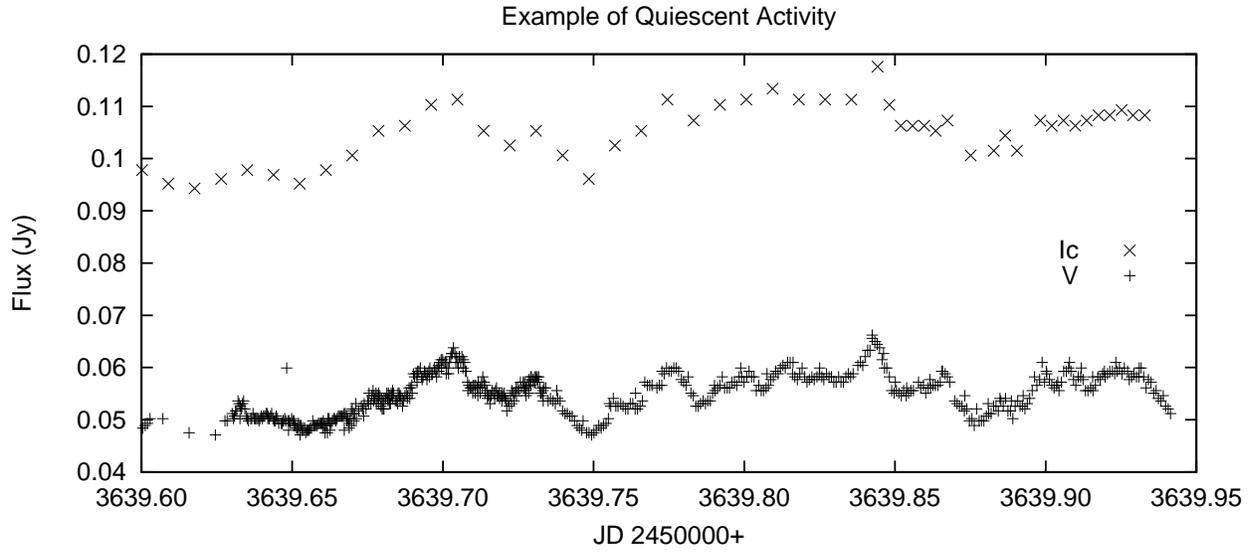}
\caption{A sample of {\it V} and {\it I$_{c}$} flickering in the photemetric light curve.}
\end{figure}

\begin{figure}
\epsscale{1}
\plotone{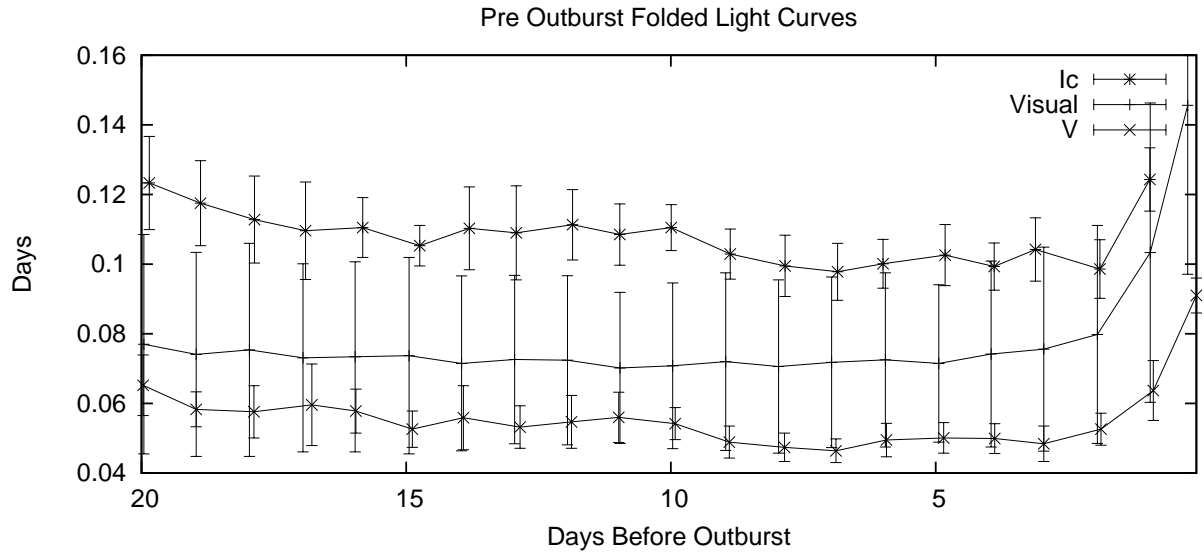}
\caption{{\it V}, {\it I$_{c}$} and visual observations averaged into 1d bins and folded at the start of an outburst.}
\end{figure}

\begin{figure}
\epsscale{1}
\plotone{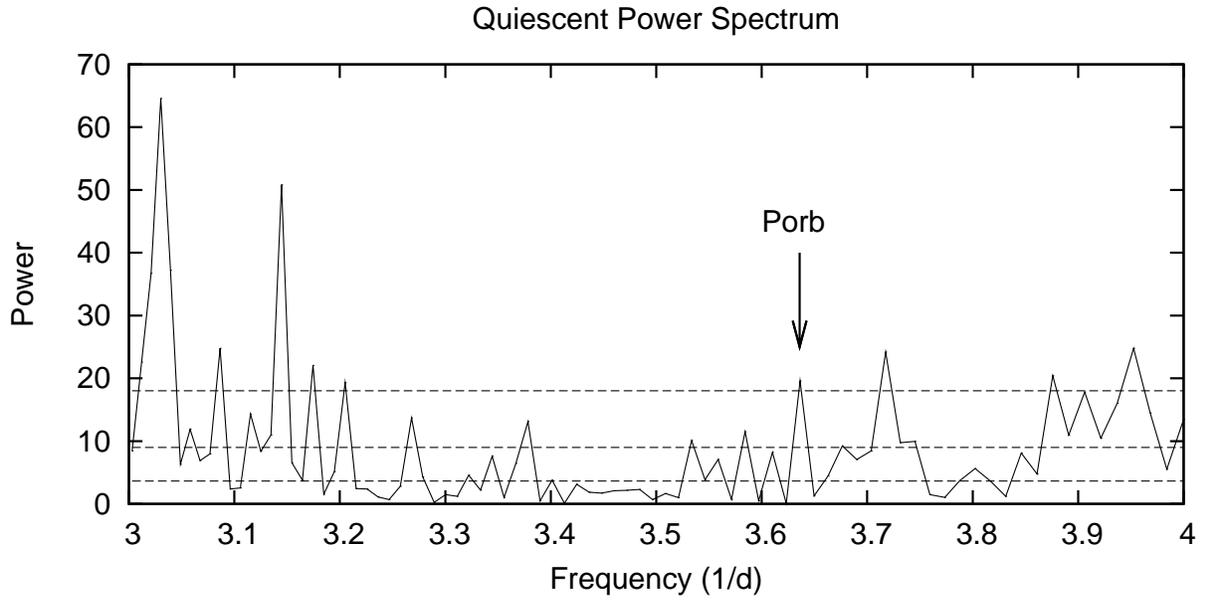}
\caption{ Red noise caused by flickering in quiescent observations. The horizontal lines represent 3, 2 and 1 sigma significance (descending). }
\end{figure}

\begin{figure}
\epsscale{1}
\plotone{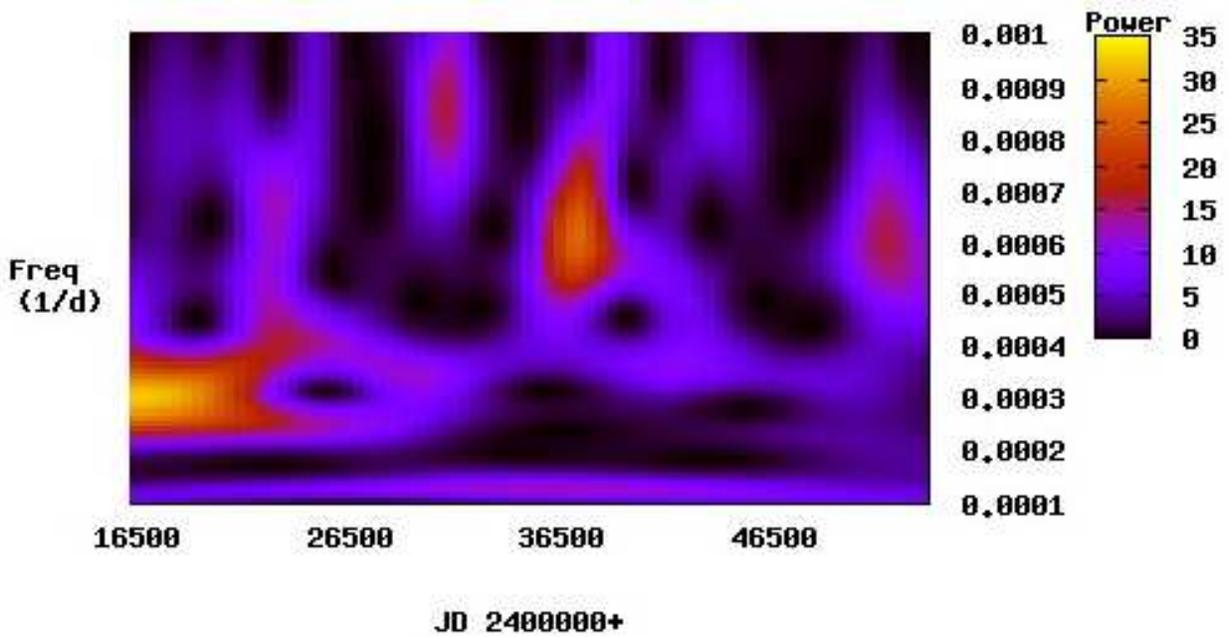}
\caption{ Wavelet analysis of 1,000-2,000d quasiperiodic activity in the AAVSO 102 year database of visual observations. }
\end{figure}

\begin{figure}
\epsscale{1}
\plotone{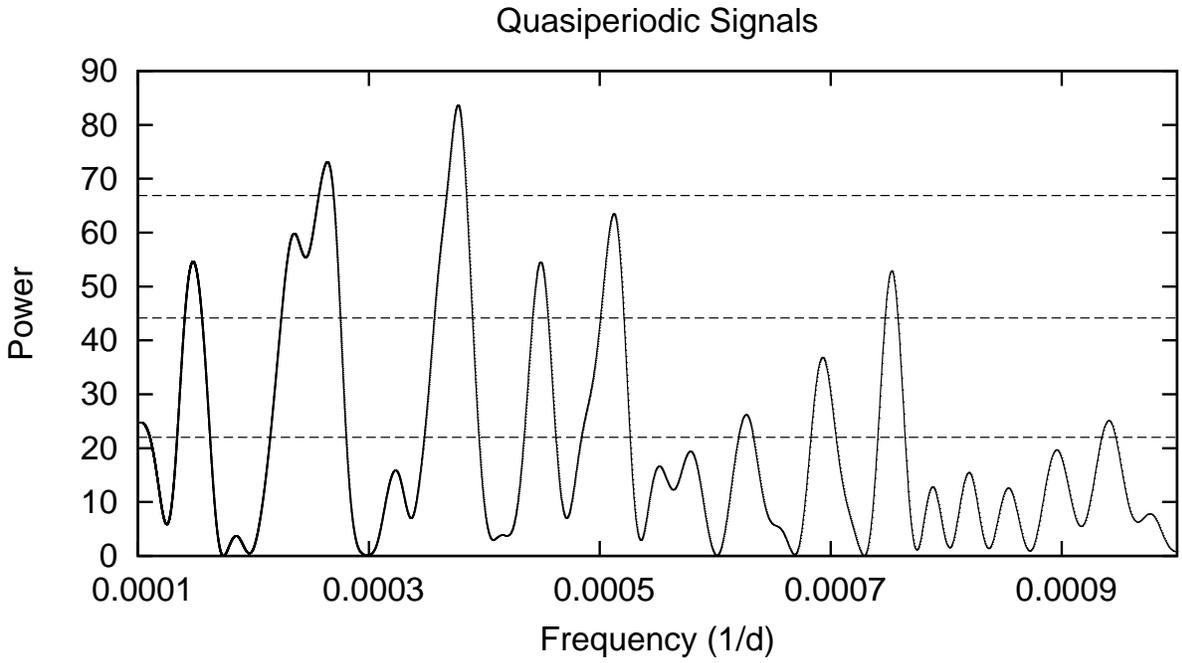}
\caption{ Fourier analysis of 1,000-2,000d quasiperiodic activity in the AAVSO 102 year database of visual observations. The horizontal lines 
represent 3, 2 and 1 sigma significance (descending).}
\end{figure}

\begin{figure}
\epsscale{1}
\plotone{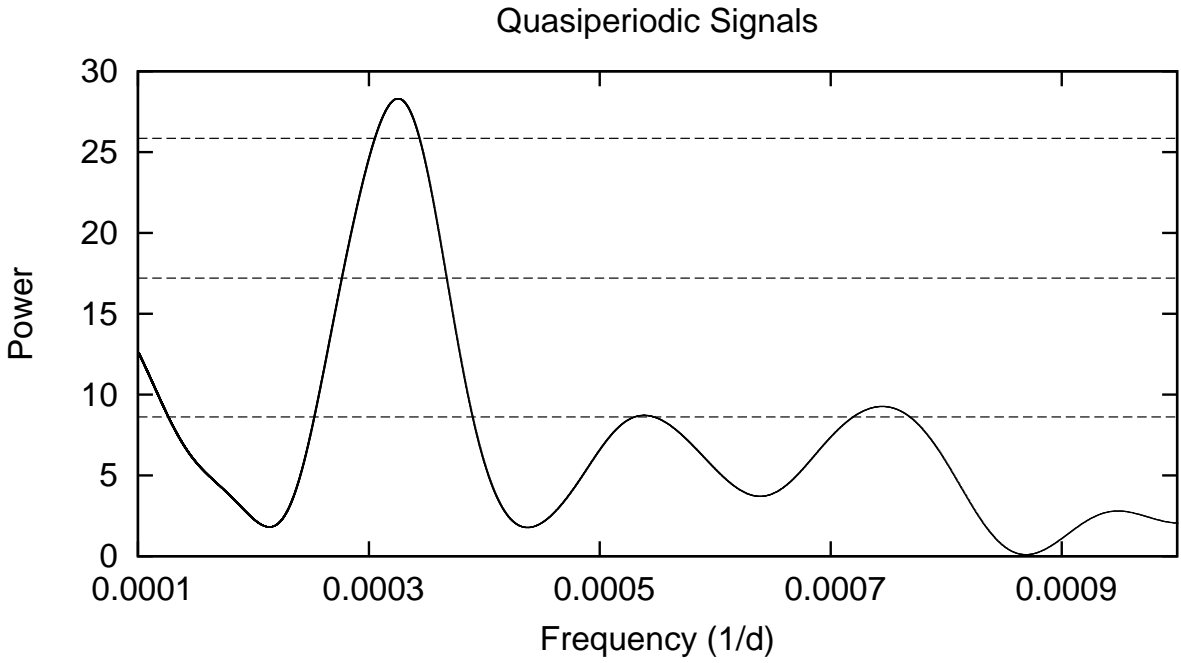}
\caption{ Fourier analysis of 1,000-2,000d quasiperiodic activity from to 2416617.7 - 2426500.0. The horizontal lines represent 3, 2 and 1 sigma 
significance (descending). }
\end{figure}

\begin{figure}
\epsscale{1}
\plotone{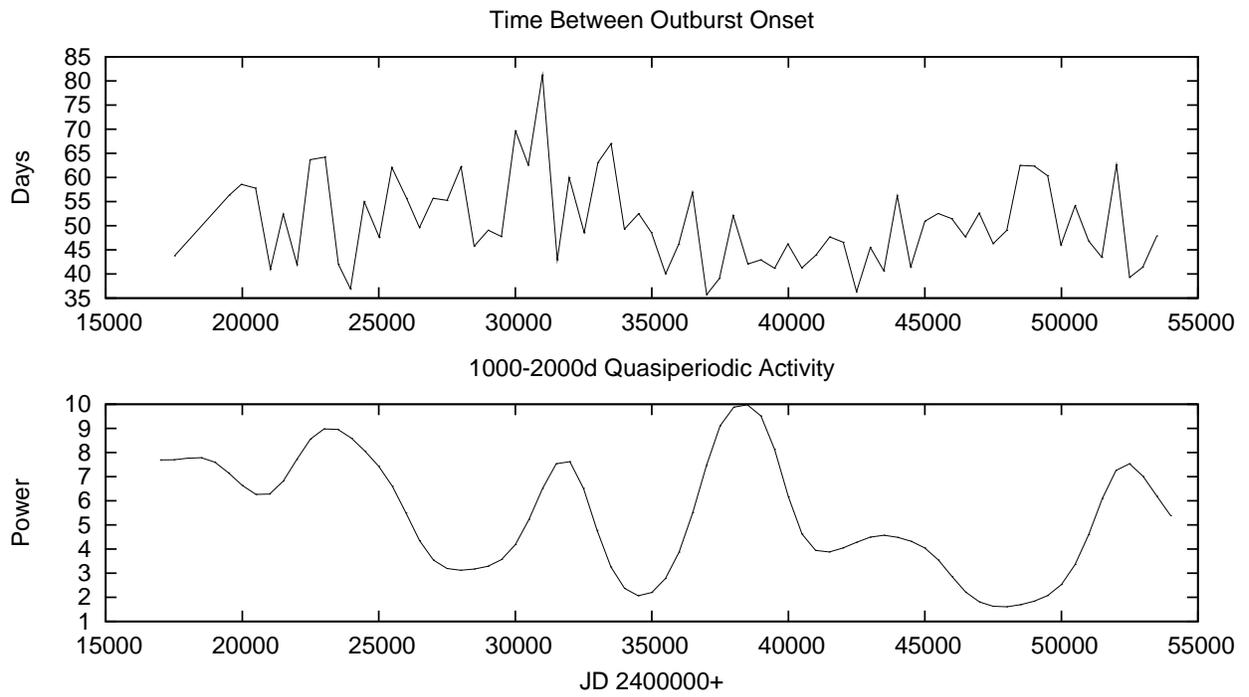}
\caption{ Average interval between outburst onsets (500d bins). A moderate bivariate
Pearson Correlation coefficient of -.356 (p$<$.01) was found between these averages and 
the 1,000-2,000d quasiperiodic activity. }
\end{figure}

\clearpage

%% the file, uncomment the \plotone, \plottwo, and
%% If you need a layout that cannot be achieved with \plotone or
%% \plottwo, you can invoke the graphicx package directly with the
%% \includegraphics command or use \plotfiddle. For more information,
%% please see the tutorial on "Using Electronic Art with AASTeX" in the
%% documentation section at the AASTeX Web site,
%% http://www.journals.uchicago.edu/AAS/AASTeX.
%%
%% The examples below also include sample markup for submission of
%% supplemental electronic materials. As always, be sure to check
%% the instructions to authors for the journal you are submitting to
%% for specific submissions guidelines as they vary from

%% This example uses \plotone to include an EPS file scaled to
%% 80% of its natural size with \epsscale. Its caption
%% has been written to indicate that additional figure parts will be
%% available in the electronic journal.

\clearpage


\begin{thebibliography}{}
\bibitem[Ak, Ozkan, \& Mattei(2001)]{Ak01}Ak, T., Ozkan, M.T., \& Mattei, J.A. 2001, A\&A, 369, 882
\bibitem[Bessell(1979)]{bes79}Bessell, M. S. 1979, \pasp, 91, 589 
\bibitem[Bianchini(1988)]{Bia88}Bianchini, A. 1988, IBVS, No. 3136 
\bibitem[Bianchini(1992)]{Bia92}Bianchini, A. Vina del Mar Workshop on Cataclysmic Variable Stars, ed. by Nikolaus Vogt, 1992. ASP Conference Series. 29, p. 284
\bibitem[Cannizzo \& Mattei(1992)]{can92}Cannizzo, J. K. \& Mattei, J. A., 1992, \apj, 401, 642 
\bibitem[Cannizzo(1993)]{can93}Cannizzo, J. K., 1993, \apj, 419, 318 
\bibitem[Cannizzo \& Mattei(1998)]{can98}Cannizzo, J. K. \& Mattei, J. A., 1998, \apj, 505, 344 
\bibitem[Collins(1999)]{Col99}Collins, P. L. 1999, J. Amer. Asoc. Var. Star Obs., 27, 1
\bibitem[Cousins(1976)]{Cou76}Cousins, A., 1976, Mem. R. Astron. Soc., 81, 25
\bibitem[Ferraz-Mello(1981)]{fer81} Ferraz-Mello, S. 1981, \aj, 86, 619
\bibitem[Foster(1996a)]{fos96a} Foster, G. 1996, \aj, 111, 541
\bibitem[Foster(1996b)]{fos96b} Foster, G. 1996, \aj, 112, 1709
\bibitem[Giovannelli(1992)]{gia92}Giovannelli, Martinez-Pais, I. G., \& Sabau Graziati, L. 1992, in ASP Conf. Ser. 29, Vina del Mar Workshop on Cataclysmic Variable Stars, 
ed. Nikolaus Vogt (San Francisco: ASP), 119
\bibitem[Hemplemann \& Kurths et al.(1990)]{hem90}Hemplemann, A., \& Kurths, J. 1990, A\&A, 232, 356
\bibitem[Honey et al.(2005)]{hon05}Honey, W. B., Bath, G. T., Charles, P. A., et al. 2005, \mnras, 236, 727
\bibitem[Hill \& Waagen(2005)]{Hil05}Hill, R. L. and Waagen, E. O. 2005, Amer. Astr. Soc.  Meeting 205, \#19.08
\bibitem[Jevtic, Mattei \& Schweitzer(2003)]{jev03} Jevtic, N., Mattei, J. A., \& Schweitzer, J. S.  2003, J. Amer. Asoc. Var. Star Obs., 31, 138
\bibitem[Kholopov(1985)]{kho85} Kholopov, P. N. 1985, General Catalog of Variable Stars, 4th edition, Moscow
\bibitem[Kiplinger et al.(1988)]{kip88} Kiplinger, A. L., Mattei, J. A., Danskin, K. H., Morgan, J. E. 1988,  J. Amer. Asoc. Var. Star 
Obs. 17, 34
\bibitem[Landolt(1992)]{lan92} Landolt, A. 1992, \aj, 104, 340
\bibitem[Osaki(1974)]{osa74} Osaki, Y. 1974, \pasj, 26, 429
\bibitem[Price et al.(2006)]{pri06} Price, A., Foster, G., Skiff, B., \& Henden, A. 2006, Amer. Astr. Soc. Meeting 209, \#162.06
\bibitem[Richman, Applegate \& Patterson(1994)]{ric94}Richman, H. R., Applegate, J. H., \& Patterson, J. 1994, \pasp, 106, 1075
\bibitem[Turner(1908)]{tur08} Turner, H. H. 1908, \mnras, 68, 544
\bibitem[Turner et al.(2006)]{Tur06} Turner, R., Price, A., Templeton, M., Waagen, E. O., \& Henden, A. 2006, Amer. Astr. Soc.  Meeting 209, \#162.05.
\bibitem[Voloshina \& Khruzina(2000)]{vol00} Voloshina, I. B., \& Khruzina, T. S. 2000, Astr. Rep., 44, 2
\end{thebibliography}
\end{document}